# Radiative thermal diode via hyperbolic metamaterials

Qizhang Li[1,2*], Haiyu He[1,3*], Qun Chen[2] and Bai Song[1,3,4†]

[1]*Beijing Innovation Center for Engineering Science and Advanced Technology, Peking University, Beijing 100871, China*

[2]*Key Laboratory for Thermal Science and Power Engineering of Ministry of Education, Department of Engineering Mechanics, Tsinghua University, Beijing 100084, China*

[3]*Department of Energy and Resources Engineering, Peking University, Beijing 100871, China*

[4]*Department of Advanced Manufacturing and Robotics, Peking University, Beijing 100871, China*

[†]Corresponding author. Email: songbai@pku.edu.cn

[*]These authors contributed equally to this work.

**ABSTRACT:** Hyperbolic metamaterials (HMMs) support propagating waves with arbitrarily large wavevectors over broad spectral ranges, and are uniquely valuable for engineering radiative thermal transport in the near field. Here, by employing a rational design approach based on the electromagnetic local density of states, we demonstrate the ability of HMMs to substantially rectify radiative heat flow. Our idea is to establish a forward-biased scenario where the two HMM-based terminals of a thermal diode feature overlapped hyperbolic bands which result in a large heat current, and suppress the reverse heat flow by creating spectrally mismatched density of states as the temperature bias is flipped. As an example, we present a few high-performance thermal diodes by pairing HMMs made of polar dielectrics and metal-to-insulator transition (MIT) materials in the form of periodic nanowire arrays, and considering three representative kinds of substrates. Upon optimization, we theoretically achieve a rectification ratio of 324 at a 100 nm gap, which remains greater than 148 for larger gap sizes up to 1 μm over a wide temperature range. The maximum rectification represents an almost 1000-fold increase compared to a bulk diode using the same materials, and is twice that of state-of-the-art designs. Our work highlights the potential of HMMs for rectifying radiative heat flow, and may find applications in advanced thermal management and energy conversion systems.

# I. INTRODUCTION

The advancement of nanotechnology over the past decades has enabled the fabrication of various subwavelength-structured optical metamaterials with a plethora of rationally designed properties unavailable from natural materials [1–3]. With a unique hyperbolic (or indefinite) dispersion, hyperbolic metamaterials (HMMs) support propagating waves with arbitrarily large wavevectors and a correspondingly large density of states, which further result in an unparalleled ability to interact with the electromagnetic near field [1–3]. Consequently, HMMs has sparked broad scientific interest and found a range of extraordinary applications including sub-diffraction imaging [4], all-angle negative refraction [5], and thermal radiation engineering [6]. Thermal radiation in the far field is limited by the Stefan-Boltzmann law for blackbody [7], which breaks down in the near field due to photon tunneling [8]. In light of their ability to support efficient tunneling for a large number of electromagnetic modes over a broad spectral range, HMMs have been pursued for enhancing radiative heat flow or even as an analog of blackbody in the near field [6,9–11], and also for improving thermophotovoltaic energy conversion devices [12–14].

In this letter, we consider HMMs for the active control of heat flow, which is of both fundamental and applied interest in energy harvesting and thermal management [15]. In particular, we consider heat flow rectification via a thermal diode [16,17]. The performance of a thermal diode is dictated by its rectification ratio, $R = \frac{(Q_F - Q_R)}{Q_R}$, where $Q_F$ and $Q_R$ are respectively the heat current under a forward and reverse temperature bias [18,19]. Thermal diode based on near-field radiation between closely spaced bodies separated by a vacuum gap can potentially reach ultrahigh rectification, especially when surface resonant modes are utilized which lead to orders-of-magnitude enhancement of $Q_F$. However, most state-of-the-art designs mediated by surface phonon polaritons (SPhPs) [20–23] and surface plasmon polaritons (SPPs) [24,25] have only predicted rectification ratios around 10, even in the extreme near field (e.g., 10 nm gap at room temperature) [20,21,23–25]. Very recently, a theoretical study employing the surface modes of thin films achieved rectifications over 140 at an experimentally accessible gap of 100 nm [26]. Compared to materials with surface modes, HMMs can support near-field heat currents of similar or even larger magnitude [6], with contributions from a much wider frequency range. However, the potential of HMMs for thermal rectification remains largely untapped.

Here, we systematically explore HMMs for thermal rectification by employing an approach proposed in our earlier work [26], with a focus on the near-field local density of states (LDOS) [27]. We theoretically show that a thermal diode with two HMM-based terminals can considerably rectify near-field radiative heat flow. Briefly, our general idea is to create a forward-biased scenario where the two terminals feature overlapped hyperbolic bands which result in a large heat current; while in the reverse scenario, the thermal transport is largely suppressed due to spectrally mismatched density of states. As an example, we rationally design a few high-performance thermal diodes by pairing HMMs made of polar dielectrics and metal-to-insulator transition (MIT) materials in the form of periodic nanowire arrays. We achieve an optimized rectification ratio of 324 at a 100 nm gap, which remains greater than 148 for larger gap sizes up to 1 μm over a wide temperature range. With recent progress in the experimental study of near-field transport between parallel planes [28–31], we expect this work to facilitate the realization of thermal diodes with unprecedented rectification ratios.

## II. BASIC CHARACTERISTICS AND NEAR-FIELD LDOS OF HMM

To analyze the LDOS characteristics of HMMs, we first introduce their basic electromagnetic responses (Fig. 1). Conceptually, HMMs are subwavelength-structured uniaxial media characterized by an effective permittivity tensor $\hat{\varepsilon} = \text{diag}(\varepsilon_\perp, \varepsilon_\perp, \varepsilon_\parallel)$, where $\varepsilon_\perp$ and $\varepsilon_\parallel$ are respectively the components perpendicular and parallel to the optical axis, with $\varepsilon_\perp \varepsilon_\parallel < 0$. The dispersion relation for *p*-polarized waves in a HMM can be written as [6]

$$\frac{\kappa^2}{\varepsilon_\parallel} + \frac{k_z^2}{\varepsilon_\perp} = \frac{\omega^2}{c^2}, \tag{1}$$

where $\kappa = \sqrt{k_x^2 + k_y^2}$ is the wavevector parallel to the HMM surface and $k_z$ is the normal wavevector inside the HMM. Equation (1) leads to hyperbolic isofrequency surfaces and permits solutions with indefinitely large $\kappa$ (Fig. 1a). The abundance of high-$\kappa$ modes further translates to an unusually large density of states in the near field. Practically, two configurations are widely used to construct HMMs. One is a multilayer with alternating metallic and dielectric films; the other consists of metallic nanowires periodically arranged in a dielectric matrix. Due to the subwavelength dimensions in these structures, the localized fields at the individual metal/dielectric interfaces can readily couple, which generates a collective response that can be viewed as from an effective homogenous medium [3].

Here, we consider HMMs in the form of a nanowire array in vacuum (Fig. 1b), the permittivity of which is obtained via the Maxwell Garnett (MG) effective medium theory [32,33] as

$$\varepsilon_\perp = \frac{(1+f)\varepsilon + (1-f)}{(1-f)\varepsilon + (1+f)}, \tag{2}$$

$$\varepsilon_\parallel = f\varepsilon + (1-f), \tag{3}$$

where $\varepsilon$ and $f$ are the permittivity and volume filling ratio of the nanowires, respectively. In general, $\varepsilon$ is complex and depends on the angular frequency $\omega$ so the criterion for HMMs should be modified as $\text{Re}(\varepsilon_\perp)\text{Re}(\varepsilon_\parallel) < 0$. This condition can be fulfilled by metals as well as materials with metal-like responses ($\text{Re}(\varepsilon) < 0$), such as polar dielectrics in the Reststrahlen band. In Fig. 1c, we plot the effective permittivity for a nanowire array of cubic boron nitride (cBN) [34,35], which is a representative polar dielectric with much potential for thermal rectification [22,26]. With a filling ratio of 0.2, we obtain two frequency bands ($\Delta_\text{I}$ and $\Delta_\text{II}$) within which the effective medium functions as a HMM. In $\Delta_\text{I}$, the medium is a Type I HMM with $\text{Re}(\varepsilon_\perp) > 0$ and $\text{Re}(\varepsilon_\parallel) < 0$, while in $\Delta_\text{II}$ it is Type II with $\text{Re}(\varepsilon_\perp) < 0$ and $\text{Re}(\varepsilon_\parallel) > 0$ (Fig. 1a, c).

With the effective permittivity, the LDOS at a distance $z$ above the medium can be calculated as [36]

$$\rho(\omega,z) = \frac{\omega^2}{2\pi^2 c^3} \left\{ \int_0^{k_0} \frac{\kappa d\kappa}{k_0 |\gamma_0|} \frac{\left(1-|r_s|^2\right)+\left(1-|r_p|^2\right)}{2} \right. \\ \left. + \int_{k_0}^\infty \frac{2\kappa^3 d\kappa}{k_0^3 |\gamma_0|} \frac{\text{Im}(r_s)+\text{Im}(r_p)}{2} e^{-2\text{Im}(\gamma_0)z} \right\}, \tag{1}$$

where $r_s$ and $r_p$ are respectively the Fresnel reflection coefficients of the $s$ and $p$ polarizations, $\gamma_0$ is the normal wavevector in vacuum, and $k_0 = \omega/c$. In Fig. 1d, we plot the LDOS of the cBN nanowire array in comparison to that of bulk cBN. Due to the dominance of SPhPs around the resonant frequency with $\text{Re}(\varepsilon_{cBN}) = -1$, bulk cBN features a sharp LDOS peak [37]. In contrast, the nanowire array has two broad LDOS peaks in the two hyperbolic bands ($\Delta_\text{I}$ and $\Delta_\text{II}$) due to contributions from the non-resonant high-$\kappa$ modes [6]. Outside the Reststrahlen band where $\text{Re}(\varepsilon_{cBN}) > 0$, the nanowire array shows an apparently lower LDOS than the bulk (Fig. 1d) because the number of contributing evanescent modes is reduced. Briefly, in the metamaterial the cutoff wavevectors are respectively $\text{Re}(\sqrt{\varepsilon_\perp})k_0$ and $\text{Re}(\sqrt{\varepsilon_\parallel})k_0$ for the $s$ and $p$ polarizations, both smaller than $\text{Re}(\sqrt{\varepsilon})k_0$ for the bulk. As bulk cBN is transformed into a nanowire array, the LDOS changes distinctly within and outside the Reststrahlen band, which proves key to realizing a large rectification.

## III. RATIONAL DESIGN OF HMM-BASED THERMAL DIODES

Inspired by the LDOS characteristics of the polar-dielectric-based HMM, we now proceed to the rational design of thermal diodes with ultrahigh rectification ratios. Our goal is to leverage the broad LDOS peaks of HMMs within the Reststrahlen band to achieve a large forward heat flux, and suppress the reverse heat flux by exploiting the much smaller LDOS elsewhere in the spectrum. Based on the theory of fluctuational electrodynamics, the radiative heat flux between two planes across a vacuum gap $d$ is given by

$$q(T_1,T_2,d) = \int_0^\infty \frac{d\omega}{4\pi^2}\left[\Theta(\omega,T_1)-\Theta(\omega,T_2)\right]\int_0^\infty d\kappa\,\kappa\left[\tau_s^{12}(\omega,\kappa)+\tau_p^{12}(\omega,\kappa)\right], \qquad (5)$$

where $\Theta(\omega,T)$ is the mean energy of a harmonic oscillator at temperature $T$ minus the zero-point contribution, $\tau_s^{12}(\omega,\kappa)$ and $\tau_p^{12}(\omega,\kappa)$ are the transmission probabilities for the *s*- and *p*-polarized modes [8]. To ensure that the MG effective medium approximation holds, the gap size should be much larger than the lattice constants of the nanowire arrays [38–40]. Considering that arrays with lattice constants on the order of 10 nm are readily attainable [41], we set the minimum gap size to be 100 nm. Moreover, since the MG theory assumes a dilute system, only small filling ratios ($f < 0.5$) are analyzed [42,43].

One way to build a HMM-based thermal diode with large rectification is to pair the polar dielectric nanowire array with an array of MIT nanowires. As an example, we consider the combination of nanowire arrays made of cBN and vanadium dioxide (VO$_2$)—a prototypical MIT material with a nominal phase transition temperature of 341 K [44–46]. The same filling ratio of 0.2 is assumed for both arrays. The terminal temperatures are set to be $T_{high}$ = 351 K and $T_{low}$ = 331 K so that VO$_2$ undergoes a complete phase transition when the temperature bias is flipped [46]. To illustrate the working mechanism of such a diode, we compare it with a diode made of bulk cBN and VO$_2$, in terms of the corresponding LDOS (Fig. 2) and the transmission probabilities (Fig. 3a-f). In the forward scenario, the VO$_2$ array is at $T_{high}$ and features a wide hyperbolic band (beyond the spectral range of Fig. 2a) with a noticeable LDOS enhancement over bulk VO$_2$, due to the metallic nature of the nanowires. In this case, the hyperbolic bands of the cBN array completely overlaps with that of the VO$_2$ array (Fig. 2a), leading to efficient photon tunneling over a wider spectral range (Fig. 3e) compared to the bulk case (Fig. 3b), and a correspondingly larger heat flux (Fig. S3 and Table S1 in the Supplementary Material [47]). In the reverse scenario, the VO$_2$ array features narrower hyperbolic bands in the spectral range outside the Reststrahlen band of cBN (Fig. 2b), where the substantially reduced LDOS of the cBN array leads to much smaller transmission (Fig. 3c, f) and heat flux (Fig. S3 and Table S1). Taken together, the HMM-based diode achieves a rectification ratio of 3 at a 100 nm gap, about 10 times larger than that of the bulk diode.

To further improve the rectification performance of the HMM-based diode, we replace the thick HMMs of cBN and VO$_2$ (Fig. 3d) with suspended thin HMM layers (Fig. 3g). Due to the reduced source volume, contribution from the broadband propagating modes is largely suppressed (Fig. 3h, i) [26,48,49]. With both layers set to be 1 μm thick, the thin-HMM-based diode yields a rectification ratio of 68, over 20 times larger than the thick-HMM case. The substantial rectification enhancement originates mainly from the dramatically reduced reverse heat flux, while the forward heat flux remains largely unaffected (Fig. S3 and Table S1). By systematically optimizing the thicknesses and filling ratios for the two nanowire layers, we obtain a maximum rectification ratio of 336, nearly 2.4 times larger than that of the thin-film diode reported in ref. [26], which employs the same base materials but is mediated by surface resonant modes.

## IV. HMM-BASED DIODES WITH REPRESENTATIVE SUBSTRATES

The thin-HMM-based thermal diode is theoretically appealing but impractical. Here, we show that comparably large rectifications can be achieved even if a substrate is added to support each of the nanowire arrays. Specifically, we demonstrate the feasibility of using three kinds of materials as the substrates, including the same materials as the nanowires, transparent dielectrics, and high-reflectivity metals (Fig. 4).

To begin with, we consider thin-film substrates using the same materials as the corresponding nanowires for ease of etching-based fabrication [41,50]. In order to provide sufficient mechanical support without introducing much undesired thermal transport, we set the substrate thickness to 10% that of the corresponding nanowire layer. After adding 100-nm-thick cBN and $VO_2$ films at the bottom of the 1-μm-thick cBN and $VO_2$ nanowire arrays, respectively, we obtain a rectification of 59 at a 100 nm gap, similar to the case with no substrates. After systematic optimization, a maximum rectification of 255 is obtained, when the thicknesses of the cBN and $VO_2$ nanowire layers are 100 nm and 7 μm, respectively. The large thickness of the $VO_2$ nanowires reflects the need to keep the $VO_2$ substrate sufficiently far away from the cBN side, which would otherwise contribute non-negligibly to the reverse heat flux. In light of this observation, we replace the $VO_2$ substrate with a transparent dielectric such as potassium bromide (KBr), which has already been used in thermal diodes for similar purposes [26]. The diode with $VO_2$ nanowires on a KBr substrate shows excellent similarity to the case without substrates both in terms of the transmission probabilities and spectral heat fluxes (Fig. S4 and Fig. 4e), and yields essentially the same rectification ratios of 65 and 324 before and after optimization, respectively. Note that the optimized rectification ratio is almost 1000 times larger than that of the bulk diode. In addition to KBr, other dielectrics with high transparency in the infrared range such as potassium chloride and calcium fluoride should also be good substrate materials.

The diodes with thin-film substrates exhibit ultrahigh thermal rectification only when they are suspended, which is still technically challenging. Alternatively, we consider metal substrates which are capable of blocking radiation from the back side and thus can be further supported on a bulk substrate (Fig. S5). Unlike the thin-film diode in ref. [26] which requires a transparent interlayer to protect the coupled surface modes, the nanowire arrays can directly sit on the metal substrates since hyperbolic modes are robust against the interfaces. By using silver (Ag) substrates for both sides of the diode, we obtain transmission probabilities very similar to the case of no substrate (Fig. 4d), despite a slight increase of the reverse heat flux (Fig. 4e). An ultrahigh rectification ratio of 174 is again obtained upon optimization.

As demonstrated above, all the diodes with substrates yield rectification ratios over 170 upon geometrical optimization at a gap size of 100 nm (Table S2). In Fig. 5a, we further plot their rectification ratios as a function of gap size together with the case of no substrate, which generally decreases with increasing gap due to the decay of contributing hyperbolic modes (Fig. S6). Among the cases with substrates, the highest rectification is consistently obtained with the transparent substrate. Impressively, when optimized at each gap size, this diode yields a maximum rectification ratio of 269 at a 500 nm gap, which remains larger than 148 even at a 1 μm gap. In addition, we show that all these diodes are expected to perform well over a wide temperature range (Fig. 5b).

# V. CONCLUSION

In summary, we have theoretically demonstrated the great potential of HMMs for rectifying near-field radiative heat flow. We first show that a HMM in the form of a polar dielectric nanowire array exhibits broad LDOS peaks within the Reststrahlen band due to a metal-like negative permittivity, but considerably smaller LDOS elsewhere with a positive permittivity. In light of these distinct LDOS characteristics, we propose a HMM-mediated rectification scheme by pairing the polar dielectric nanowire array with an array of MIT nanowires. The hyperbolic bands of these two HMMs readily overlap when the MIT nanowires are in the metallic phase, but are completely separated upon phase transition, leading to a large contrast in the radiative heat fluxes. As an example, our diodes consisting of optimized cBN and $VO_2$ nanowire arrays supported on various substrates can in theory achieve a maximum rectification ratio of 324 at a 100 nm gap which is nearly 1000 times larger than that of the corresponding bulk diode. These diodes remain highly effective for larger gap sizes up to 1 μm over a wide temperature range. Our results offer a new perspective for manipulating thermal radiation with HMMs, and further highlight the efficacy of our LDOS-based rational design approach.

# ACKNOWLEDGEMENT

This work was supported by the National Natural Science Foundation of China (Grant No. 52076002 and No. 51836004), the Beijing Innovation Center for Engineering Science and Advanced Technology (BIC-ESAT), the XPLORER PRIZE from the Tencent Foundation, and the High-performance Computing Platform of Peking University.

# REFERENCES

[1] A. Poddubny, I. Iorsh, P. Belov, and Y. Kivshar, Nat. Photonics **7**, 948 (2013).

[2] P. Shekhar, J. Atkinson, and Z. Jacob, Nano Converg. **1**, 14 (2014).

[3] L. Ferrari, C. Wu, D. Lepage, X. Zhang, and Z. Liu, Prog. Quantum Electron. **40**, 1 (2015).

[4] Z. Liu, H. Lee, Y. Xiong, C. Sun, and X. Zhang, Science **315**, 1686 (2007).

[5] T. Xu, A. Agrawal, M. Abashin, K. J. Chau, and H. J. Lezec, Nature **497**, 470 (2013).

[6] S.-A. Biehs, M. Tschikin, and P. Ben-Abdallah, Phys. Rev. Lett. **109**, 104301 (2012).

[7] G. Chen, *Nanoscale Energy Transport and Conversion: A Parallel Treatment of Electrons, Molecules, Phonons, and Photons* (Oxford Univ. Press, 2005).

[8] B. Song, A. Fiorino, E. Meyhofer, and P. Reddy, AIP Adv. **5**, 53503 (2015).

[9] Y. Guo, C. L. Cortes, S. Molesky, and Z. Jacob, Appl. Phys. Lett. **101**, 131106 (2012).

[10] S.-A. Biehs, M. Tschikin, R. Messina, and P. Ben-Abdallah, Appl. Phys. Lett. **102**, 131106 (2013).

[11] X. Liu, R. Z. Zhang, and Z. Zhang, ACS Photonics **1**, 785 (2014).

[12] J.-Y. Chang, Y. Yang, and L. Wang, Int. J. Heat Mass Transf. **87**, 237 (2015).

[13] S. Jin, M. Lim, S. S. Lee, and B. J. Lee, Opt. Express **24**, A635 (2016).

[14] N. Vongsoasup, M. Francoeur, and K. Hanamura, Int. J. Heat Mass Transf. **115**, 326 (2017).

[15] Y. Li, W. Li, T. Han, X. Zheng, J. Li, B. Li, S. Fan, and C.-W. Qiu, Nat. Rev. Mater. **6**, 488 (2021).

[16] N. Li, J. Ren, L. Wang, G. Zhang, P. Hänggi, and B. Li, Rev. Mod. Phys. **84**, 1045 (2012).

[17] G. Wehmeyer, T. Yabuki, C. Monachon, J. Wu, and C. Dames, Appl. Phys. Rev. **4**, 41304

(2017).

[18]  C. W. Chang, D. Okawa, A. Majumdar, and A. Zettl, Science **314**, 1121 (2006).

[19]  C. R. Otey, W. T. Lau, and S. Fan, Phys. Rev. Lett. **104**, 154301 (2010).

[20]  J. Huang, Q. Li, Z. Zheng, and Y. Xuan, Int. J. Heat Mass Transf. **67**, 575 (2013).

[21]  L. P. Wang and Z. M. Zhang, Nanoscale Microscale Thermophys. Eng. **17**, 337 (2013).

[22]  A. Ghanekar, J. Ji, and Y. Zheng, Appl. Phys. Lett. **109**, 123106 (2016).

[23]  L. Tang and M. Francoeur, Opt. Express **25**, A1043 (2017).

[24]  G. Xu, J. Sun, H. Mao, and T. Pan, J. Appl. Phys. **124**, 183104 (2018).

[25]  J. Shen, X. Liu, H. He, W. Wu, and B. Liu, J. Quant. Spectrosc. Radiat. Transf. **211**, 1 (2018).

[26]  Q. Li, H. He, Q. Chen, and B. Song, arXiv:2103.04035 (2021).

[27]  A. V Shchegrov, K. Joulain, R. Carminati, and J.-J. Greffet, Phys. Rev. Lett. **85**, 1548 (2000).

[28]  B. Song, D. Thompson, A. Fiorino, Y. Ganjeh, P. Reddy, and E. Meyhofer, Nat. Nanotechnol. **11**, 509 (2016).

[29]  A. Fiorino, D. Thompson, L. Zhu, B. Song, P. Reddy, and E. Meyhofer, Nano Lett. **18**, 3711 (2018).

[30]  M. Ghashami, H. Geng, T. Kim, N. Iacopino, S. K. Cho, and K. Park, Phys. Rev. Lett. **120**, 175901 (2018).

[31]  J. DeSutter, L. Tang, and M. Francoeur, Nat. Nanotechnol. **14**, 751 (2019).

[32]  J. C. M. Garnett and J. Larmor, Philos. Trans. R. Soc. London. Ser. A, Contain. Pap. a Math. or Phys. Character **205**, 237 (1906).

[33]  X. J. Wang, J. L. Abell, Y. P. Zhao, and Z. M. Zhang, Appl. Opt. **51**, 1521 (2012).

[34] A. Narayanaswamy and G. Chen, Appl. Phys. Lett. **82**, 3544 (2003).

[35] K. Chen, B. Song, N. K. Ravichandran, Q. Zheng, X. Chen, H. Lee, H. Sun, S. Li, G. A. G. Udalamatta Gamage, F. Tian, Z. Ding, Q. Song, A. Rai, H. Wu, P. Koirala, A. J. Schmidt, K. Watanabe, B. Lv, Z. Ren, L. Shi, D. G. Cahill, T. Taniguchi, D. Broido, and G. Chen, Science **367**, 555 (2020).

[36] K. Joulain, J.-P. Mulet, F. Marquier, R. Carminati, and J.-J. Greffet, Surf. Sci. Rep. **57**, 59 (2005).

[37] S.-A. Biehs, E. Rousseau, and J.-J. Greffet, Phys. Rev. Lett. **105**, 234301 (2010).

[38] X. Liu and Z. Zhang, ACS Photonics **2**, 1320 (2015).

[39] S.-A. Biehs, P. Ben-Abdallah, F. S. S. Rosa, K. Joulain, and J.-J. Greffet, Opt. Express **19**, A1088 (2011).

[40] V. Fernández-Hurtado, F. J. García-Vidal, S. Fan, and J. C. Cuevas, Phys. Rev. Lett. **118**, 203901 (2017).

[41] Z. Huang, X. Zhang, M. Reiche, L. Liu, W. Lee, T. Shimizu, S. Senz, and U. Gösele, Nano Lett. **8**, 3046 (2008).

[42] K. J. Button, *Infrared and Millimeter Waves*, Vol. 13 (Academic, 1985).

[43] J. Y. Chang, P. Sabbaghi, and L. Wang, Int. J. Heat Mass Transf. **158**, 120023 (2020).

[44] J. Barker, H. Verleur, and H. Guggenheim, Phys. Rev. Lett. **17**, 1286 (1966).

[45] H. Paik, J. A. Moyer, T. Spila, J. W. Tashman, J. A. Mundy, E. Freeman, N. Shukla, J. M. Lapano, R. Engel-Herbert, W. Zander, J. Schubert, D. A. Muller, S. Datta, P. Schiffer, and D. G. Schlom, Appl. Phys. Lett. **107**, 163101 (2015).

[46] S. Lee, K. Hippalgaonkar, F. Yang, J. Hong, C. Ko, J. Suh, K. Liu, K. Wang, J. J. Urban, X. Zhang, C. Dames, S. A. Hartnoll, O. Delaire, and J. Wu, Science **355**, 371 (2017).

[47] See Supplementary Material for radiative characteristics of the VO$_2$ nanowire array with very small filling ratios, performance of the HMM-based diodes as a function of filling ratio, spectral heat fluxes of different diodes (two bulks, two thick HMMs and two thin HMMs), transmission probabilities of the HMM-based diodes with same-material and transparent substrates, performance of the HMM-based diodes with Ag substrates sitting on two representative bulk materials, gap-dependent rectification ratios and the percentage of heat flux contributed from the hyperbolic modes.

[48] S.-A. Biehs, D. Reddig, and M. Holthaus, Eur. Phys. J. B **55**, 237 (2007).

[49] M. Francoeur, M. P. Mengüç, and R. Vaillon, Appl. Phys. Lett. **93**, 43109 (2008).

[50] Z. Huang, H. Fang, and J. Zhu, Adv. Mater. **19**, 744 (2007).

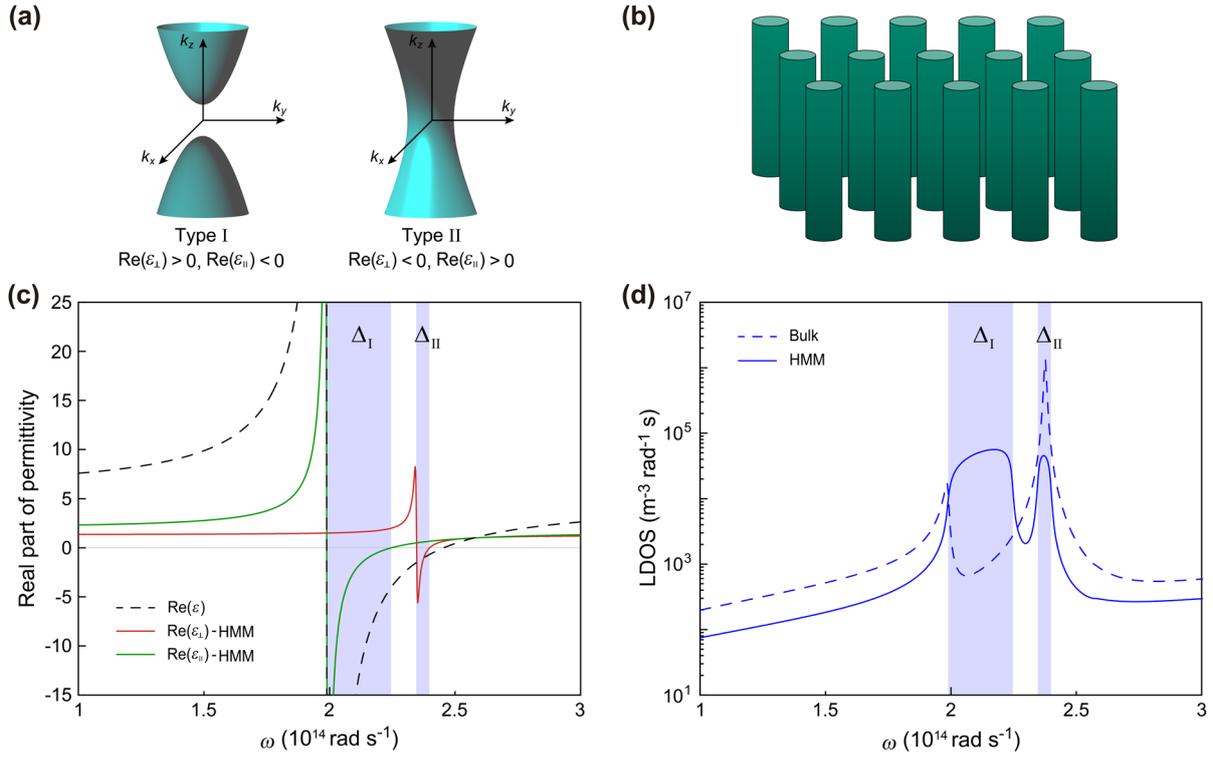

**Fig. 1.** Basic characteristics of HMMs. (a) Illustration of the hyperbolic isofrequency surfaces for a Type I and a Type II HMM. (b) Schematic of a HMM consisting of a periodical array of nanowires in vacuum. (c) The real part of the cBN permittivity ($\varepsilon$) and the two components of the effective permittivity for a cBN nanowire array ($\varepsilon_\perp$ and $\varepsilon_\parallel$) with a filling ratio of 0.2. (d) The LDOS at $z = $ 100 nm above bulk cBN and the cBN nanowire array. The blue shades $\Delta_\mathrm{I}$ and $\Delta_\mathrm{II}$ mark the Type I and Type II hyperbolic bands, respectively.

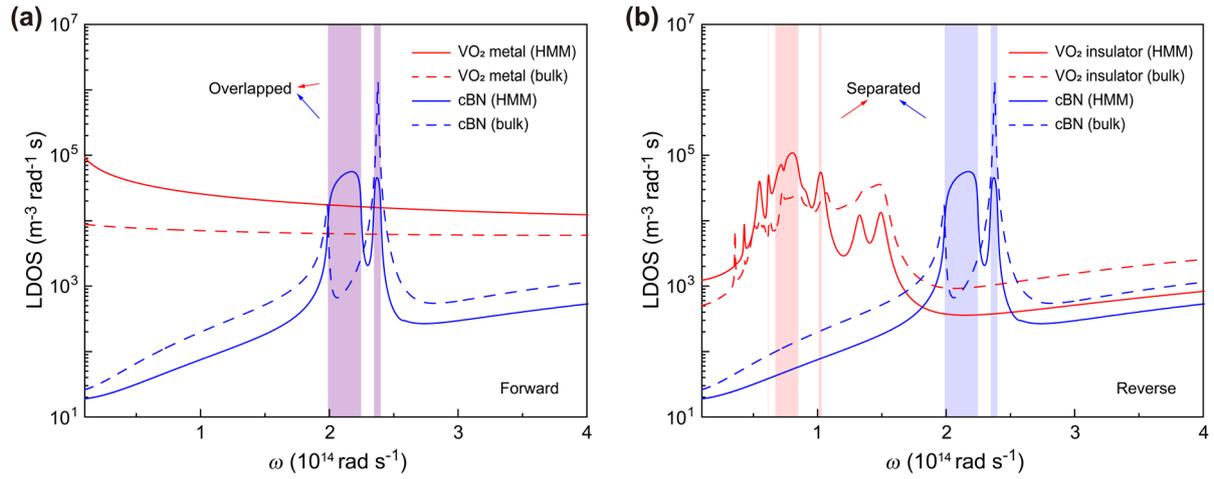

**Fig. 2.** Rectification mechanism of a HMM-based thermal diode consisting of a cBN nanowire array and a VO$_2$ nanowire array, illustrated in terms of the LDOS 100 nm above the media. (a) In the forward-biased scenario, VO$_2$ is a metal and the hyperbolic bands of the two HMMs overlap. (b) Under a reverse bias, VO$_2$ is an insulator and the hyperbolic bands of the two HMMs are spectrally separated. A filling ratio of 0.2 is assumed for all the arrays. The LDOS for the cBN side is a replot of Fig. 1d for ease of comparison.

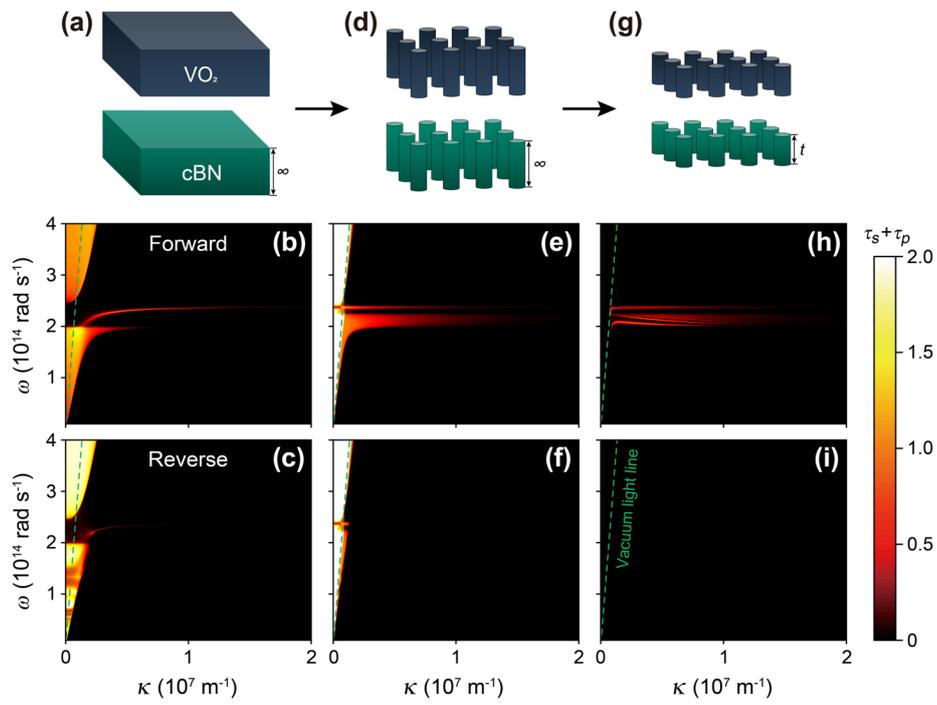

**Fig. 3.** The key for large rectification using HMM-based diodes revealed by a comparison of their transmission probabilities with that of a bulk diode. (a-c) bulk diode, (d-f) two thick HMMs, (g-i) two thin HMMs (1 μm thick). The vacuum gap is 100 nm wide, and the same filling ratio of 0.2 is assumed for all the nanowire arrays.

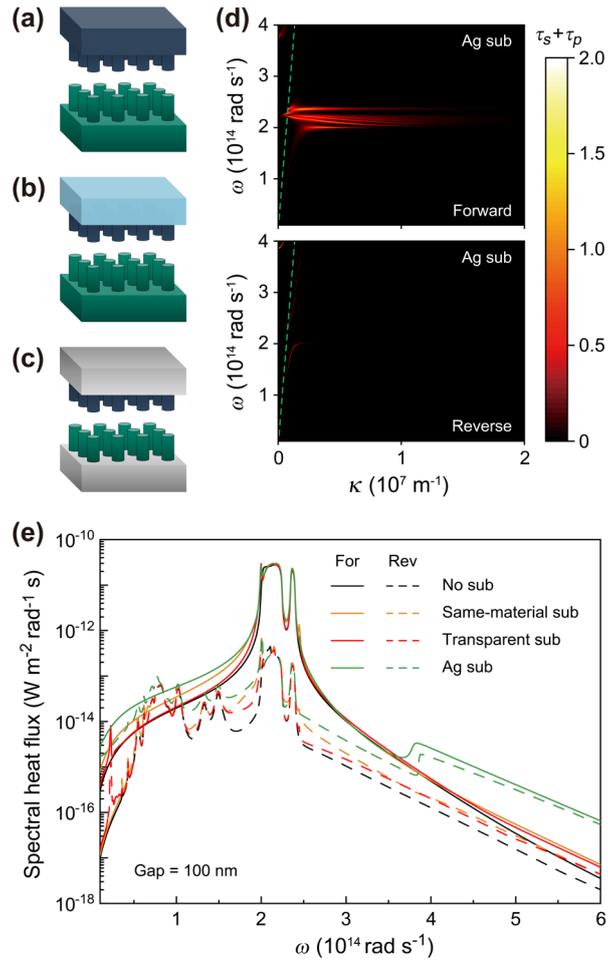

**Fig. 4.** Analysis of HMM-based diodes with different kinds of substrates. (a) The substrates employ the same materials as the corresponding HMMs. (b) The cBN nanowires sit on a cBN substrate while the $VO_2$ nanowires on a transparent substrate (KBr). (c) Ag substrates for both HMMs. (d) Transmission probabilities for the diode with Ag substrates. (e) Spectral heat fluxes in the forward (solid) and reverse (dashed) scenarios for different diodes. The vacuum gap is 100 nm wide, and all the thin HMM layers are 1 μm thick with a filling ratio of 0.2. The Ag substrates are assumed to be semi-infinite.

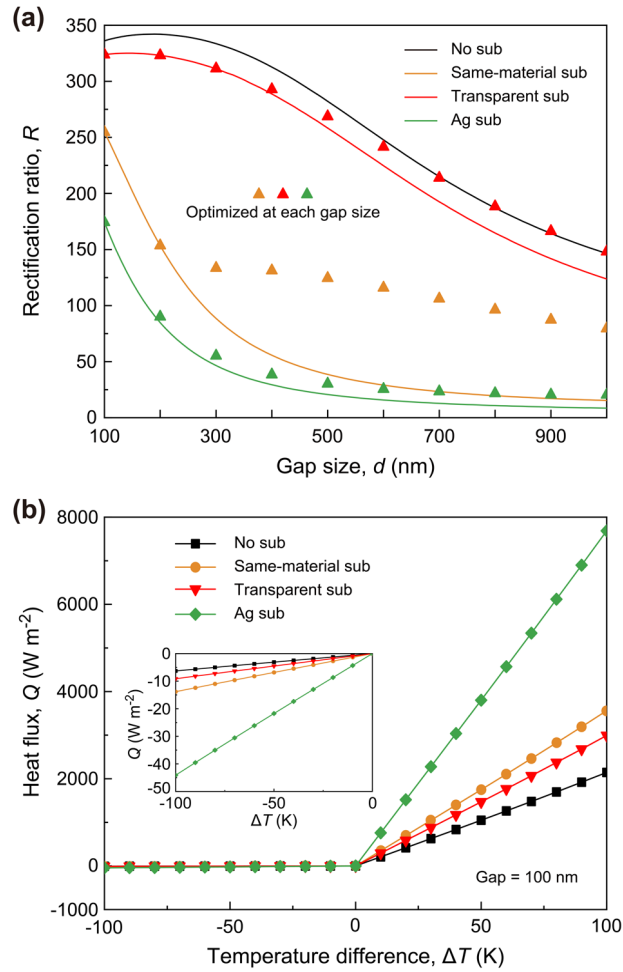

**Fig. 5.** Performance of HMM-based diodes after optimization of the filling ratio and the thicknesses of the nanowire layers. (a) Rectification ratios as a function of gap size. The lines are for diodes optimized only at 100 nm gap, while the markers indicate values after optimization at the corresponding gap sizes. (b) Heat fluxes under both forward and reverse temperature biases at a gap of 100 nm. The temperatures for the VO$_2$ and the cBN side are set as $T_1 = 341 + \Delta T$ and $T_2 = 341 - \Delta T$, respectively. In the optimization, the minimum thicknesses of the nanowire layers are set to be 100 nm [43].